\documentclass{emulateapj}
\shorttitle{Interior angular momentum of core hydrogen burning stars}

\shortauthors{C.~Aerts et al.}

\begin{document} 

\title{The interior angular momentum of core hydrogen burning stars from
  gravity-mode oscillations}

\author{C.~Aerts\altaffilmark{1,2,3}, T. Van Reeth\altaffilmark{1,3},
A. Tkachenko\altaffilmark{1}}

\altaffiltext{1}{Institute of Astronomy, KU\,Leuven, Celestijnenlaan 200D,
B-3001 Leuven, Belgium}
\altaffiltext{2}{Department of Astrophysics, IMAPP, Radboud University
Nijmegen, P.O.\ Box 9010, 6500 GL Nijmegen, The Netherlands}
\altaffiltext{3}{Kavli Institute for Theoretical Physics, University of California, Santa
Barbara, CA 93106, USA}
%\altaffiltext{6}{??}

\begin{abstract}
  A major uncertainty in the theory of stellar evolution is the angular momentum
  distribution inside stars and its change during stellar life.  We compose a
  sample of 67 stars in the core-hydrogen burning phase with a $\log\,g$ value
  from high-resolution spectroscopy, as well as an asteroseismic estimate of the
  near-core rotation rate derived from gravity-mode oscillations detected in
  space photometry. This assembly includes 8 B-type stars and 59 AF-type stars,
  covering a mass range from 1.4 to 5\,M$_\odot$, i.e., it concerns
  intermediate-mass stars born with a well-developed convective core.  The
    sample covers projected surface rotation velocities
    $v\sin\,i \in[9,242]\,$km\,s$^{-1}$ and core rotation rates up to $26\mu$Hz,
    which corresponds to 50\% of the critical rotation frequency. We find
    deviations from rigid rotation to be moderate in the single stars of this
    sample.  We place the near-core rotation rates in an evolutionary context
  and find that the core rotation must drop drastically before or during the
  short phase between the end of the core-hydrogen burning and the onset of
  core-helium burning. We compute the spin parameter, which is the ratio of
  twice the rotation rate to the mode frequency (also known as the inverse
  Rossby number), for 1682 gravity modes and find the majority (95\%) to occur
  in the sub-inertial regime. The ten stars with Rossby modes have spin
  parameters between 14 and 30, while the gravito-inertial modes cover the range
  from 1 to 15.
\end{abstract}


\section{Introduction}

Rotation has a significant impact on the life of a star
\citep[e.g.,][]{Maeder2009}. It induces a multitude of hydrodynamical processes
that affect stellar structure but have remained poorly calibrated by data. While
surface abundances and surface rotation give some constraints for stellar
models, one needs calibrations of the interior rotation and chemical mixing
profiles of stars to evaluate the theoretical concepts that have been developed
to describe transport phenomena and their interplay
\citep[e.g.,][]{Heger2000,Meynet2013}. Here, we are concerned with the interior
rotation rate of stars in the core-hydrogen burning phase of their life (aka the
main sequence).

Asteroseismology based on long-term high-precision monitoring of nonradial
oscillations is the best way to deduce the interior rotation of stars, following
the case of helioseismology \citep[e.g.][]{Thompson2003}.  Early estimates of
the interior rotation of main sequence stars were achieved from rotational
splitting of a few pressure modes in $\beta\,$Cep stars from ground-based data
\citep{Aerts2003,Pamyatnykh2004,Briquet2007}. However, it concerned only very
rough estimates of the ratio of the near-core to envelope rotation rates, with
values from 1 to $\sim 4$ for three stars with a mass between 8 and
10\,M$_\odot$ \citep[][for a summary]{Aerts2008}.

A major breakthrough in the derivation of interior rotation rates was achieved
thanks to the NASA {\it Kepler\/} mission, once it had delivered light curves
with a duration longer than two years. The early and so far majority of interior
rotation rates were obtained from the rotational splitting of dipole mixed modes
in low- and intermediate-mass evolved stars \citep[subgiants and red giants with
a mass between 0.8 and
3.3\,M$_\odot$;][]{Beck2012,Mosser2012,Deheuvels2012,Deheuvels2014,Deheuvels2015}.
These seismic results constituted an unanticipated result, since the cyclic
rotation frequency $\Omega/2\pi$ of the stellar core (denoted as
$\Omega_{\rm core}$ hereafter) of these evolved stars turned out to be two
orders of magnitude lower than theoretical predictions. This pointed to stronger
coupling between the stellar core and envelope during and/or after the main
sequence, irrespective of the star having undergone a helium flash or not. This
major shortcoming of the theory of angular momentum transport remains unsolved
\citep[e.g.][]{Tayar2013,Cantiello2014,Eggenberger2017}.   In Sect.\,2,
we shed new light on this topic by considering the interior rotation rates
derived so far from {\it Kepler\/} and BRITE photometry of red-giant
progenitors, i.e., intermediate-mass gravity-mode pulsators on the main
sequence.

In Sect.\,3, we provide the observational information to evaluate the
  assumptions made in the theory of angular momentum transport by waves.  A
  promising physical ingredient that can explain the observed behavior of the
  core-to-surface rotation of main sequence stars is the transport of angular
  momentum by low-frequency internal gravity waves created at the interface
  between the convective core and the radiative envelope, observed from space
  photometry \citep{AertsRogers2015,Rogers2015}.  The theory of the interaction
  between such waves and (differential) rotation inside stars requires proper
  treatment of the various forces at play. Based on our findings for the
  rotation, we provide the spin parameters (inverse Rossby numbers) of the
  gravity modes as input for angular momentum computations. We end the paper
  with a discussion in Sect.\,4.  

\section{Core rotation from gravity-mode oscillations}

Gravity-mode oscillations of main sequence stars have periods from 0.5 to 3\,d
\citep[e.g.,][Chapter\,2]{Aerts2010}. Recent space photometry revealed period
spacing patterns of such modes in B-, A- and F-type stars covering the entire
main sequence \citep[e.g.][]{Degroote2010,Papics2014,Kurtz2014,Saio2015,
  VanReeth2015,Murphy2016,Ouazzani2017,Papics2017,Zwintz2017}, just as predicted
by theory \citep{Miglio2008,Bouabid2013}.  Such modes have their dominant mode
energy in the near-core region of the star and are therefore excellent probes of
the chemical gradient left behind during the main sequence shrinkage of the
convective core.  The detected gravity modes have rotational kernels that are
typically 5 to 10 times larger in a narrow near-core region than in the extended
radiative envelope \citep[e.g.,][their Figs\,2]{Triana2015,VanReeth2016}.

For stars that reveal both gravity and pressure modes with rotational splitting,
derivation of $\Omega_{\rm core}$ and $\Omega_{\rm env}$ is possible in an
almost model-independent way.  This was first achieved by \citet{Kurtz2014} for
the A-type {\it Kepler\/} target KIC\,11145123 and soon followed by
\citet{Saio2015} for the F-type star KIC\,9244992.  More so-called hybrid
pulsators with rotational splitting in both types of modes have been found
meanwhile. 

The majority of intermediate-mass gravity-mode pulsators found in the {\it
  Kepler\/} data reveal quantitative information on the near-core rotation but
not on the envelope rotation.  As derived by \citet{VanReeth2016} and
\citet{Ouazzani2017}, the slope of the period spacing pattern of prograde,
zonal, or retrograde modes of consecutive radial order delivers a direct probe
of $\Omega_{\rm core}$, even in the absence of rotational splitting.  A value of
$\Omega_{\rm env}$ can be deduced from rotational splitting of pressure
modes. In the absence of identified pressure modes, the surface rotation can
still be derived with high precision from the detection of rotational modulation
when this phenomenon occurs together with gravity modes in the light curves.
Otherwise, only a lower limit for $\Omega_{\rm env}$ can be deduced from a
spectroscopic measurement of the projected surface rotation velocity
$\Omega_{\rm env}\,R\sin\,i$ \citep{Murphy2016,VanReeth2016}. This can be
achieved by adopting a reasonable value for the stellar radius while paying
attention to line-profile broadening induced by all the tangential velocity
fields due to gravity modes \citep{Aerts2014a}.

\begin{figure*}
\begin{center}
\rotatebox{270}{\resizebox{13cm}{!}{\includegraphics{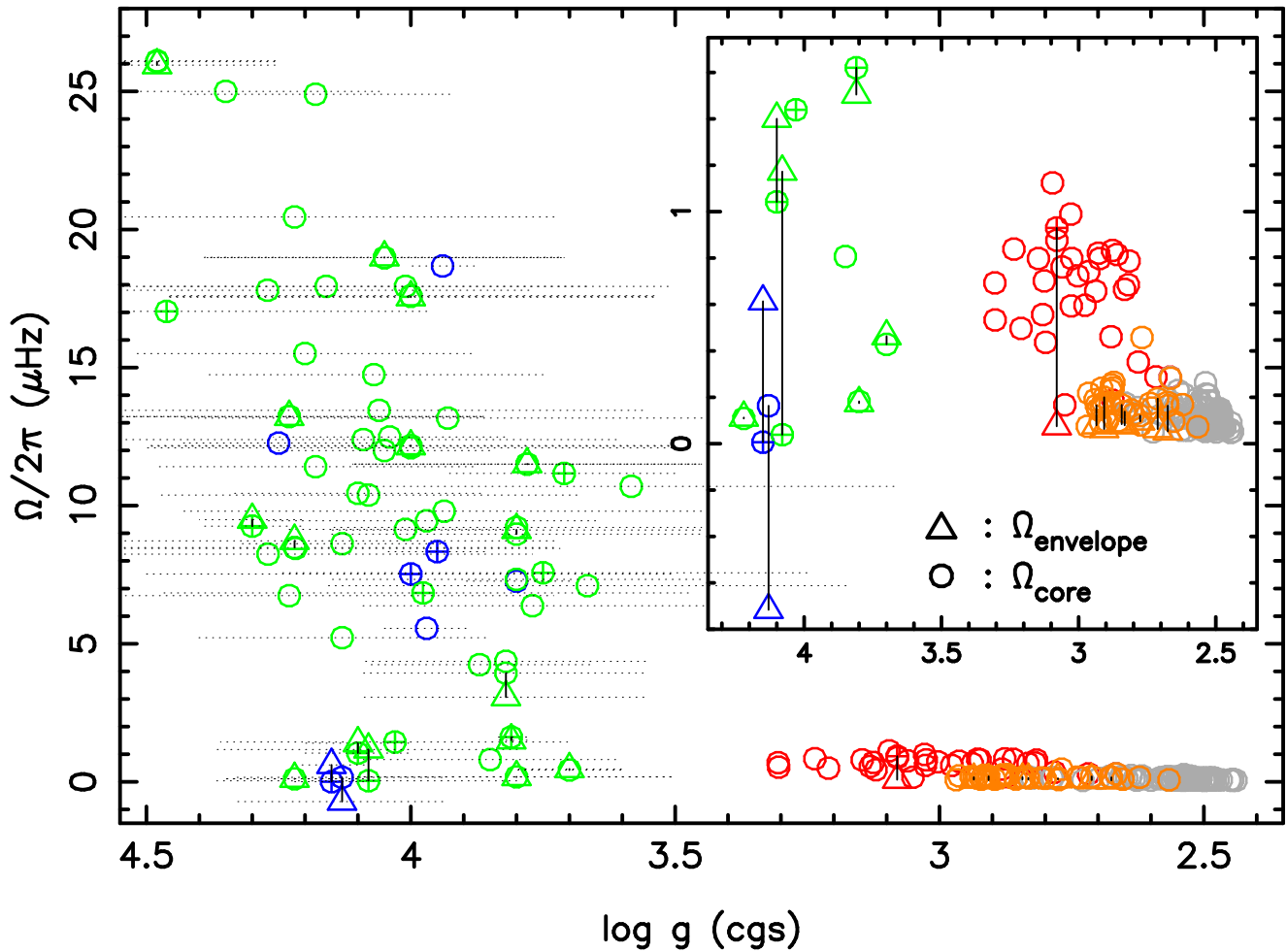}}}
\end{center}
\caption{Core rotation rates (circles) as a function of
  spectroscopically-derived gravity for core-hydrogen burning stars with a mass
  between 1.4 and 2.0\,M$_\odot$ (green) and 3 to $\sim\,5$\,M$_\odot$ (blue)
  derived from dipole prograde gravito-inertial modes
in main sequence stars.  Surface rotation rates
  (triangles) are deduced from pressure modes or from rotational
  modulation. Errors on the rotation rates are smaller than the symbol size,
  while the errors on the gravity are indicated by dotted lines.
  Asteroseismically derived rotation rates and gravities for evolved stars with
  solar-like oscillations in the mass range $1.4\,$M$_\odot<$M$<3.3\,$M$_\odot$
  have been added (errors smaller than symbol sizes for both quantities): RGB
  stars (red), red clump stars (gray) and secondary clump stars (orange).
  Binary components are indicated with an extra $+$ sign inside the circles. The
  errors of $\log\,g$ have been omitted in the zoomed window for clarity.}
\label{rotation}
\end{figure*}

Figure\,\ref{rotation} includes 67 main sequence stars from \citet{Kurtz2014,
  Saio2015, Triana2015, Moravveji2016, Murphy2016, VanReeth2016,
  SchmidAerts2016, Ouazzani2017,Papics2017, Sowicka2017,Kallinger2017}, and Guo
et al.\ (submitted), covering spectral types from mid B to early F.  All
  these studies rely on {\it Kepler\/} or BRITE space photometry.  Their
gravity modes revealed $\Omega_{\rm core}$ through asteroseismology and their
$\log\,g$ resulted from high-resolution spectroscopy (except for the four stars
in \citet{Ouazzani2017}, for which we took $\log\,g$ from the {\it Kepler\/}
input catalogue and estimated its error from \citet{VanReeth2015}, their
Fig.\,2). For 18 of the stars, the surface rotation rate is also available,
either from asteroseismology of pressure modes or from rotational
modulation. Twelve main sequence stars are member of a spectroscopic binary.

The 59 green stars in Fig.\,\ref{rotation} cover
  $1.4\,$M$_\odot<$M$<2.0\,$M$_\odot$ and $v\sin\,i\in [9,170]\,$km\,s$^{-1}$.
  According to \citet{ZorecRoyer2012}, stars in this mass range rotate on
  average with $v\sin\,i=148\,$km\,s$^{-1}$, with a dispersion of
  $54\,$km\,s$^{-1}$. Given that none of the $v\sin\,i$ studies of large
  ensembles take into account pulsational line broadening, while this
  overestimates $v\sin\,i$ \citep{Aerts2014a}, our sample of AF stars is
  representative in terms of surface rotation for this mass range.  On the other
  hand, our sample contains only 8 B stars with a mass in
  $3\,$M$_\odot<$M$<5\,$M$_\odot$. While they cover a broad range of
  $v\sin\,i\in [18,242]\,$km\,s$^{-1}$, they are not representative of the
  bimodal distribution for this mass range \citep[Table\,4
  in][]{ZorecRoyer2012}.  For both the B and AF stars in Fig.\,\ref{rotation},
  the highest core rotation frequency corresponds to $\sim\,50\%$ of the
  critical value in the Roche formalism. At present, core rotation from gravity
  modes for stars rotating faster than half critical are not available from
  seismic modeling due to lack of mode identification.

We deduce from Fig.\,\ref{rotation} that the deviation from rigid rotation in
our sample of 67 stars is low to moderate, pointing to strong core-to-envelope
coupling during core hydrogen burning, irrespective of the value of the rotation
rate. Unfortunately, the errors for the spectroscopic $\log\,g$ are too
  large for this quantity to serve as a proxy of the evolutionary stage of the
  sample stars.  Ideally, we want to use the central hydrogen fraction $X_c$
  from seismic modeling on the $x-$axis rather than $\log\,g$.  At present this
  time-consuming task was only done for two B stars, revealing both to be in the
  first part of the main sequence with $X_c=0.63\pm0.01$ and 0.50$\pm 0.01$, in
  agreement with the spectroscopic $\log\,g$ of the former being 0.2\,dex higher
  than the one of the latter \citep{Moravveji2015,Moravveji2016}. For five of
  the eight F stars in the inset in Fig.\,\ref{rotation}, a seismic estimate of
  the evolutionary stage is available from comparison of the morphology of the
  gravity-mode period spacings: $X_c\in [0.01,0.22]$.

Differences in the core-to-surface rotation, even if moderate, are
  significant for several stars in Fig.\,\ref{rotation}. Both faster cores than
  surfaces and vice versa occur.  Large nonrigidity occurs in some close
binaries, where tidal forces are active. Even though we cannot pinpoint
  $X_c$ values for all stars at present, the results in Fig.\,\ref{rotation}
  require an angular momentum redistribution mechanism that can explain the
  diversity of the measured core-to-envelope rotation during core hydrogen
  burning. One such mechanism is discussed in the following section. 

To interpret the results in Fig.\,\ref{rotation} in a more global
evolutionary context, we also show the core and surface rotation rates of
successors of these 67 stars, i.e., red giants on the RGB (red), in the red
clump (gray), and in the secondary clump (orange), limiting to the 152 stars
with masses above 1.4\,M$_\odot$ from \citet{Mosser2012,Deheuvels2015} and one
binary from \citet{Beck2014}.  For all these evolved stars, the gravity in the
figure is derived from scaling relations of solar-like oscillations (errors
  are smaller than the symbol size; B.\ Mosser kindly made the masses and radii
  of the stars in \citet{Mosser2012} available to us).  As argued by
\citet{Mosser2014}, stars born with M$>2.1\,$M$_\odot$ avoid a helium flash
after the hydrogen shell burning. Their core contraction during hydrogen shell
burning occurs on a short timescale (the so-called Hertzsprung gap) and the
onset of core helium burning happens quietly in non-degenerate matter. These
stars are hard to catch in their shell burning phase but reveal themselves as
secondary clump stars (mass range up to 3.3\,M$_\odot$ in
Fig.\,\ref{rotation}). On the other hand, the successors of the stars born with
$1.4\,$M$_\odot<$M$<2.1\,$M$_\odot$ can readily be observed while burning
hydrogen in a shell surrounding their helium core and start helium burning
violently in degenerate matter.

The importance of the stellar mass and the occurrence (or not) of a convective
core and/or $\mu-$gradient zone for the evolution of the angular momentum was
already emphasized by \citet{Tayar2013} and \citet{Eggenberger2017}.  This
becomes even more prominent from our sample of intermediate-mass gravity-mode
pulsators: efficient slow-down of the stellar core happens already on the main
sequence for stars with a convective core.  The three slowest rotators among the
B stars have masses between 3.0 and 3.3\,M$_\odot$ and are thus progenitors of
the most massive orange stars; two of these are mid main sequence while the one
with the counter-rotating envelope is young with $X_c=0.63$.

\section{Angular momentum transport by waves}

2D simulations of angular momentum transport by internal gravity waves of a
  3\,M$_\odot$ star at birth are able to explain the observed core-to-envelope
  rotation in Fig.\,\ref{rotation} in a qualitative sense \citep{Rogers2015}.
  Such simulations are computationally too demanding to be coupled to stellar
  evolution computations. The latter thus must resort to analytical treatments
  of angular momentum transport. These rely on approximations \citep[e.g.,][for
  a didactic chapter]{Mathis2011}.

Observed sub- and super-inertial waves can only propagate in a mode cavity
set by the Br\"untt-V\"ais\"al\"a frequency $N$. Nonlinear wave interactions for
the low-frequency regime can at first instance be ignored because the mode
frequencies (denoted as $\nu$) fulfill $\nu<N$.  The displacement vector of the
observed coherent heat-driven gravity modes excited in the partial ionization
zone of iron-like elements are dominantly horizontal.  Indeed, multicolor
photometry and high-precision line-profile variations in spectroscopy reveal a
typical ratio of 10 to 100 between the horizontal and vertical component of the
displacement vector of the gravity modes \citep{DeCatAerts2002}. There is no
reason why this would not be the case for convectively driven internal gravity
waves. This property simplifies the theoretical treatment of the wave transport
\citep{Mathis2011}.  The adiabatic and Cowling approximations are appropriate to
describe the low-frequency gravity waves in the near-core region of our sample
stars.

\begin{figure}
\begin{center}
\rotatebox{270}{\resizebox{6.5cm}{!}{\includegraphics{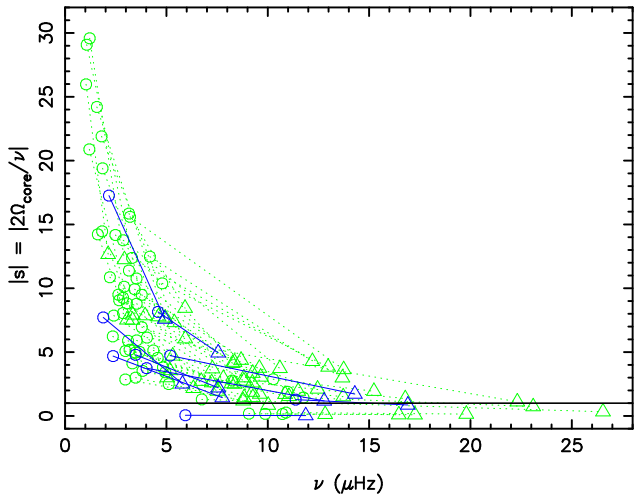}}}
\end{center}
\caption{Spin parameters for 47 main sequence stars in Fig.\,1 for which we
  identified the mode wavenumbers (same color convention). For each star,
  the range in frequencies of detected mode series of consecutive radial order
  is indicated by a line, from the circle that represents the mode with the
  lowest frequency in the corotating frame to the triangle representing the mode
  with the highest frequency. For all the F-type stars with $|s|>15$, the high
  spin parameters are caused by low-frequency retrograde Rossby modes.}
\label{spin}
\end{figure}

Ignoring the Lorentz and tidal forces, this
brings us to the key parameter to evaluate the approximations that can or
cannot be made in the treatment of the angular momentum transport by waves: the
interior rotation frequency of the star with respect to the wave frequencies.
For the stars on the main sequence indicated in blue and green in
Fig.\,\ref{rotation}, we provide an estimate of the importance of the Coriolis
force for their transport of angular momentum. This importance is expressed by
the spin parameter, also known as the inverse of the Rossby number and defined
here as $s=2\,\Omega_{\rm core}/\nu$, where $\nu$ is the cyclic frequency of the
gravity wave under study in a reference frame of the star corotating with
$\Omega_{\rm core}$. Defining a corotating reference frame is not obvious
in the case of nonrigid rotation; we choose to define $s$ with respect to the
value of $\Omega (r)/2\pi$ that is most robustly determined. Following the properties
of the rotational kernels of the gravity modes, this is the value in the
near-core $\mu$-gradient region \citep{Triana2015,VanReeth2016}.

In Fig.\,\ref{spin} we show the computed spin parameters based on the
asteroseismic $\Omega_{\rm core}$ and on the frequencies of the identified
gravity-mode series with consecutive radial order for the 7 B-type stars in
\citet{Papics2014,Papics2015,Papics2017} and 40 F-type stars in
\citet{VanReeth2015}, all of which are included in Fig.\,\ref{rotation}. For
each star, we connect the lowest- and highest-frequency mode (in the corotating
frame) to visualize the range of $s$ for each series in each star and compare it
to a Rossby number of 1 (horizontal black line).  Figure\,\ref{spin} represents
1682 modes in these 47 stars, all of which have been detected and identified
from 4-year {\it Kepler\/} space photometry.  We find that most of the stars
have gravito-inertial modes rather than pure gravity modes: 1607 modes have
frequencies in the sub-inertial regime. Among the detected modes are 273 (17\%)
retrograde Rossby modes occuring in 10 F-type stars in the sample.  The slowest
rotators reveal 75 (i.e., $<5\%$) pure gravity modes in the super-inertial
regime.  About 12\% of the modes in the 7 B stars and 4\% of the modes in the 40
AF stars occur in the super-inertial regime.

We conclude that only the modes in the slowest rotators fulfil the formal
mathematical condition on the spin parameter to apply the so-called Traditional
Approximation (TA) for the Coriolis force \citep{Townsend2003,Mathis2011}.  The
detected Rossby modes have by far the highest spin parameters, covering
$s\in [14,30]$, while the gravito-inertial modes have $s\in [1,15]$.  We
  point out that the core rotation rates in Fig.\,\ref{rotation} were obtained
  by adopting the TA and assuming rigid rotation for the observed
  gravito-inertial modes of the stars. The assumption of rigidity is not a
  drawback, because these modes have by far their largest mode energy in the
  near-core region while the mode kernels and mass distribution in the stellar
  envelope hardly contribute to the core rotation values \citep[Figs 12 and 2
  in][respectively]{Kurtz2014,VanReeth2016}. On the other hand, it was shown by
  \citet[][Fig.\,1]{Ouazzani2017}, who considered both 1D and 2D treatments of
  the modes for static equilibrium models, that the TA still provides a good
approximation in the case of zonal and prograde gravito-inertial modes for the
frequency regimes and spin parameters treated here and used to derive
  Fig.\,\ref{rotation}. However, the TA is less justified for the 20 retrograde
gravito-inertial modes among the 1607 sub-inertial modes.  Modelling the 293
retrograde gravito-inertial and Rossby modes will benefit from a 2D
non-perturbative treatment of the rotation as developed by \citet{Ouazzani2012}.

\section{Discussion}

An important aspect to consider along with
simulations of angular momentum transport by internal gravity waves 
is the chemical mixing induced by these waves and its comparison with
mixing prescriptions adopted in 1D stellar evolution codes. Hints that
pulsational mixing might be dominant over rotational mixing were found from a
sample of pulsating magnetic and non-magnetic OB-type stars \citep{Aerts2014b}.
A quantitative evaluation of the level of mixing in the near-core region and in
the radiative envelope can be obtained from gravity-mode
asteroseismology \citep[e.g.,][]{Moravveji2015,Moravveji2016}. This opens new
perspectives to include seismically calibrated prescriptions for mixing and
angular momentum transport based on multi-D numerical simulations of 
waves in stellar evolution computations. First
attempts to compute chemical mixing due to gravity waves look promising when
compared with asteroseismic results \citep{RogersMcElwaine2017}.

More gravity-mode pulsators than those treated here are under study, but they
either have too few unambiguously identified modes to derive $\Omega_{\rm core}$
\citep[e.g.,][]{Zwintz2017} or they have not been observed in spectroscopy to
derive their $v\sin\,i$ and $\log\,g$ \citep[e.g.,][]{Ouazzani2017}. Including
all those with identified modes but without a spectroscopic $\log\,g$ in
Fig.\,\ref{rotation} requires forward modeling to derive a seismic $X_c$ and
$\log\,g$.  Progress to populate Fig.\,\ref{rotation} more densely from deeper
exploitation of the {\it Kepler\/} and BRITE data and to transform this figure
into a true ``evolutionary'' diagram is expected in the near future.

Unfortunately, we currently lack $\Omega_{\rm core}$ from gravity modes for
stars born with $>5\,$M$_\odot$, and in particular of core-collapse
supernova progenitors.  A major breakthrough is expected from large samples of
OB-type stars in the Milky Way and in the Large Magellanic Cloud to be observed
during almost one year with the TESS mission \citep[TESS Southern CVZ;][launch
in 2018]{Ricker2016} and for the long pointings of the PLATO mission
\citep[][launch foreseen in 2026]{Rauer2014}. Only after deducing the interior
rotation of single and binary OB-type stars from asteroseismology
covering the hydrogen and helium core and shell burning stages will we be able
to understand angular momentum evolution and confront it with the one of white
dwarfs and neutron stars.

\acknowledgments CA and TVR are grateful for the kind hospitality and
opportunity to perform part of this research at the Kavli Institute of
Theoretical Physics, University of California at Santa Barbara, USA.  The
research leading to these results has received funding from the European
Research Council (ERC) under the European Union's Horizon 2020 research and
innovation programme (grant agreement N$^\circ$670519: MAMSIE), from the
Research Foundation Flanders (FWO, grant agreements G.0B69.13 and V4.272.17N)
and from the National Science Foundation of the United States under Grant NSF
PHY11--25915.
\\[0.2cm]

\end{document}